\documentclass[aps,prc,preprint,superscriptaddress,showpacs,floatfix]{revtex4}
\usepackage{graphicx}

\begin{document}


\title{Shell structure underlying the evolution of
quadrupole collectivity in $^{38}$S and $^{40}$S
probed by transient-field $g$-factor measurements on fast
radioactive beams.}




\author{A.E.~Stuchbery}
\affiliation{Department of Nuclear Physics, The Australian National
University, Canberra, ACT 0200, Australia}
\author{A.D.~Davies}
\affiliation{National Superconducting Cyclotron Laboratory, Michigan
State University, East Lansing, MI 48824 USA}
\affiliation{Department of Physics and Astronomy, Michigan State
University, East Lansing, MI 48824 USA}
\author{P.F.~Mantica}
\affiliation{National Superconducting Cyclotron Laboratory, Michigan
State University, East Lansing, MI 48824 USA}
\affiliation{Department of Chemistry, Michigan State University,
East Lansing, MI 48824 USA}
\author{P.M.~Davidson}
\affiliation{Department of Nuclear Physics, The Australian National
University, Canberra, ACT 0200, Australia}
\author{A.N.~Wilson}
\affiliation{Department of Nuclear Physics, The Australian National
University, Canberra, ACT 0200, Australia} \affiliation{Department
of Physics, The Australian National University, Canberra, ACT 0200,
Australia}
\author{A.~Becerril}
\affiliation{National Superconducting Cyclotron Laboratory, Michigan
State University, East Lansing, MI 48824 USA}
\affiliation{Department of Physics and Astronomy, Michigan State
University, East Lansing, MI 48824 USA}
\author{B.A.~Brown}
\affiliation{National Superconducting Cyclotron Laboratory, Michigan
State University, East Lansing, MI 48824 USA}
\affiliation{Department of Physics and Astronomy, Michigan State
University, East Lansing, MI 48824 USA}
\author{C.M.~Campbell}
\affiliation{National Superconducting Cyclotron Laboratory, Michigan
State University, East Lansing, MI 48824 USA}
\affiliation{Department of Physics and Astronomy, Michigan State
University, East Lansing, MI 48824 USA}
\author{J.M.~Cook}
\affiliation{National Superconducting Cyclotron Laboratory, Michigan
State University, East Lansing, MI 48824 USA}
\affiliation{Department of Physics and Astronomy, Michigan State
University, East Lansing, MI 48824 USA}
\author{D.C.~Dinca}
\affiliation{National Superconducting Cyclotron Laboratory, Michigan
State University, East Lansing, MI 48824 USA}
\affiliation{Department of Physics and Astronomy, Michigan State
University, East Lansing, MI 48824 USA}
\author{A.~Gade}
\affiliation{National Superconducting Cyclotron Laboratory, Michigan
State University, East Lansing, MI 48824 USA}
\author{S.N.~Liddick}
\affiliation{National Superconducting Cyclotron Laboratory, Michigan
State University, East Lansing, MI 48824 USA}
\affiliation{Department of Chemistry, Michigan State University,
East Lansing, MI 48824 USA}
\author{T.J.~Mertzimekis}
\affiliation{National Superconducting Cyclotron Laboratory, Michigan
State University, East Lansing, MI 48824 USA}
\author{W.F.~Mueller}
\affiliation{National Superconducting Cyclotron Laboratory, Michigan
State University, East Lansing, MI 48824 USA}
\author{J.R.~Terry}
\affiliation{National Superconducting Cyclotron Laboratory, Michigan
State University, East Lansing, MI 48824 USA}
\affiliation{Department of Physics and Astronomy, Michigan State
University, East Lansing, MI 48824 USA}
\author{B.E.~Tomlin}
\affiliation{National Superconducting Cyclotron Laboratory, Michigan
State University, East Lansing, MI 48824 USA}
\affiliation{Department of Chemistry, Michigan State University,
East Lansing, MI 48824 USA}
\author{K.~Yoneda}
\affiliation{National Superconducting Cyclotron Laboratory, Michigan
State University, East Lansing, MI 48824 USA}
\author{H.~Zwahlen}
\affiliation{National Superconducting Cyclotron Laboratory, Michigan
State University, East Lansing, MI 48824 USA}
\affiliation{Department of Physics and Astronomy, Michigan State
University, East Lansing, MI 48824 USA}


\date{\today}

\begin{abstract}
The shell structure underlying shape changes in neutron-rich nuclei
between $N=20$ and $N=28$ has been investigated by a novel
application of the transient field technique to measure the
first-excited state $g$~factors in $^{38}$S and $^{40}$S produced as
fast radioactive beams. Details of the new methodology are
presented. In both $^{38}$S and $^{40}$S there is a fine balance
between the proton and neutron contributions to the magnetic
moments. Shell model calculations which describe the level schemes
and quadrupole properties of these nuclei also give a satisfactory
explanation of the $g$~factors. In $^{38}$S the $g$~factor is
extremely sensitive to the occupation of the neutron $p_{3/2}$ orbit
above the $N=28$ shell gap as occupation of this orbit strongly
affects the proton configuration. The $g$~factor of deformed
$^{40}$S does not resemble that of a conventional collective nucleus
because spin contributions are more important than usual.
\end{abstract}



\maketitle


\section{Introduction}

\begin{figure}
\resizebox{80mm}{!}{\includegraphics{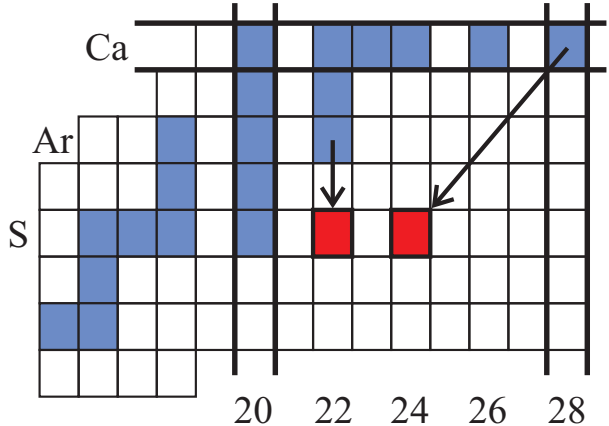}} \caption{(Color
online) Region of the nuclear chart indicating the primary beam
isotopes $^{40}$Ar and $^{48}$Ca used to produce the $^{38}$S and
$^{40}$S secondary beams (red). The $N=20,28$ closed shells are
shown outlined in black. Stable isotopes are shaded (blue).}
\label{fig:chart}
\end{figure}

In a recent Letter \cite{Dav2006} we presented the first application
of a high-velocity transient-field (HVTF) technique \cite{Stuc2004a}
to measure the $g$~factors of excited states of neutron-rich nuclei
produced as fast radioactive beams. Questions on the nature and
origins of deformation between $N=20$ and $N=28$ were addressed by
measuring the $g$ factors of the 2$_1^+$ states in
$^{38}_{16}$S$_{22}$ and $^{40}_{16}$S$_{24}$. Fig.~\ref{fig:chart}
shows the relevant part of the nuclear chart and indicates the
primary beams used to produce the isotopes of interest. In the
present paper we give a more comprehensive description of the
experiment and discuss the interpretation in greater detail. A more
technically-oriented discussion of the new technique will be
published elsewhere \cite{Stu2006a}.

\begin{figure}
\resizebox{80mm}{!}{\includegraphics{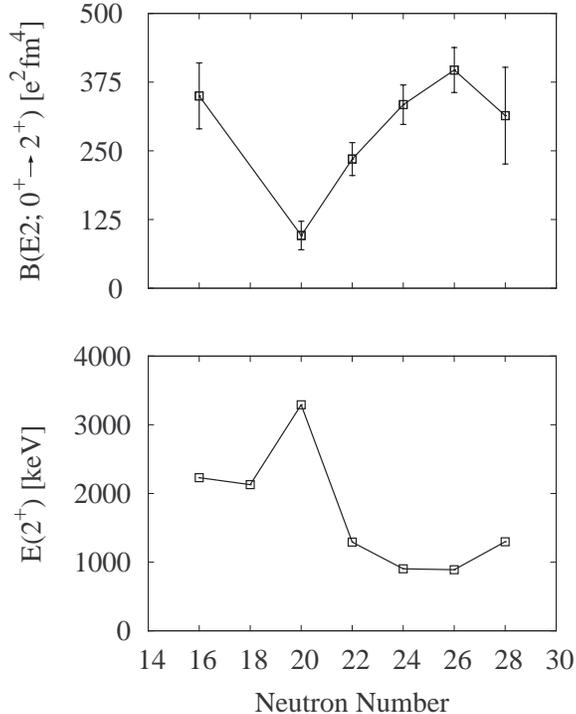}}
\caption{Previously-measured $E(2^+)$ and $B(E2)$ values for the
sulfur isotopes \protect \cite{Raman}. } \label{fig:E2BE2}
\end{figure}

Figure~\ref{fig:E2BE2} shows the 2$^+_1$ excitation energies and the
$B(E2;0^+_1 \rightarrow 2^+_1)$ values for the sulfur isotopes. It
is apparent from the reduction in the excitation energies of the
2$^+_1$ states and the increase in $B(E2)$ values that the neutron
rich isotopes between $N=20$ and $N=28$ develop collective features.
Furthermore, the $N=28$ nucleus $^{44}_{16}$S$_{28}$ does not have
the high 2$^+_1$ energy and small $B(E2)$ like the closed-shell
$N=20$ isotope $^{36}_{16}$S$_{20}$. The data in
Fig.~\ref{fig:E2BE2} together with additional information on the
low-excitation level structures imply that the even sulfur isotopes
between $N=20$ and $N=28$ undergo a transition from spherical at
$^{36}_{16}$S$_{20}$, to prolate deformed in $^{40}_{16}$S$_{24}$
and $^{42}_{16}$S$_{26}$, and that the $N=28$ nucleus
$^{44}_{16}$S$_{28}$ appears to exhibit collectivity of a
vibrational character~\cite{Sche1996,Glasm1997,Wing2001,Sohl2002}.
However the evolution of collective features in these nuclei has
underlying causes that remain unclear. Some have argued that a
weakening of the $N=28$ shell gap is important
\cite{Sohl2002,Wern1994}, while others have opined that the effect
of adding \textit{neutrons} to the $f_{7/2}$ orbit is primarily to
reduce the \textit{proton} $s_{1/2}$-$d_{3/2}$ gap and that a
weakening of the $N=28$ shell gap is not needed to explain the
observed collectivity near $^{44}$S \cite{Cott1998}. There have been
several theoretical studies discussing the erosion of the $N=28$
shell closure and the onset of deformation (e.g.
Refs.~\cite{rod02,cau04} and references therein). To help resolve
questions on the nature and origins of deformation between $N=20$
and $N=28$, we have used a novel technique to measure the $g$
factors of the 2$_1^+$ states in $^{38}_{16}$S$_{22}$ and
$^{40}_{16}$S$_{24}$.

The remainder of the paper is arranged as follows:
Section~\ref{sect:expt} describes the experimental procedures
including radioactive beam production, targets and apparatus. The
experimental results and details of the data analysis are given in
Section~\ref{sect:results}. Our shell model calculations are
presented and discussed in Section~\ref{sect:discussion}.

\section{Experimental procedures} \label{sect:expt}

\subsection{Overview}

The experiments reported here use transient fields to study
neutron-rich nuclei produced as radioactive beams by fast
fragmentation. In our approach the nuclear states of interest are
excited and aligned by intermediate-energy Coulomb excitation
\cite{Glasm1998}. The nucleus is then subjected to the transient
field in a higher velocity regime than has been used previously for
moment measurements, which causes the nuclear spin to precess.
Finally, the nuclear precession angle, to which the $g$~factor is
proportional, is observed via the perturbed $\gamma$-ray angular
correlation measured using a multi-detector array. It is important
to note that the transient-field technique has sensitivity to the
{\em sign} of the $g$~factor, which in itself can be a
distinguishing characteristic of the state under study, since the
signs of the spin contributions to the proton and the neutron
$g$~factors are opposite.


The transient field (TF) is a velocity-dependent magnetic hyperfine
interaction experienced by the nucleus of a swift ion as it
traverses a magnetized ferromagnetic material
\cite{Kol1980,Spe2002}. For light ions ($Z \lesssim 16$) the maximum
TF strength is reached when $v = Zv_0$, i.e. the ion velocity $v$
matches the $K$-shell electron velocity $Zv_0$ ($v_0 = c/137$, Bohr
velocity). Since the transient field arises from polarized electrons
carried by the moving ion its strength falls off as the ion velocity
exceeds $Zv_0$ and the ion becomes fully stripped; a transient-field
interaction will not occur for fast radioactive beams with energies
near 100 MeV/nucleon until most of that energy is removed. Thus
slowing the fast fragment beams to velocities where the transient
field can be effective is an essential feature of the measurement.

\subsection{Radioactive beam production and properties}

The experiment was conducted at the Coupled Cyclotron Facility of
the National Superconducting Cyclotron Laboratory at Michigan State
University.  Secondary beams of $^{38}$S and $^{40}$S were produced
from 140 MeV/nucleon primary beams of $^{40}$Ar and $^{48}$Ca,
respectively, directed onto a $\sim 1$~g/cm$^2$ $^9$Be fragmentation
target at the entrance of the A1900 fragment separator
\cite{Morr2003}. An acrylic wedge degrader 971 mg/cm$^2$ thick and a
0.5\% momentum slit at the dispersive image of the A1900 were
employed.  The wedge degrader allowed the production of highly pure
beams and also reduced the secondary beam energy to $\sim
40$~MeV/nucleon. Further details of the radioactive beams are given
in Table~\ref{tab:beams}. The $^{38}$S ($^{40}$S) measurement ran
for 81 (68) hours. The 40 MeV/nucleon beams were made incident upon
a target consisting of contiguous layers of Au and Fe, 355 mg/cm$^2$
thick and 110 mg/cm$^2$ thick, respectively.

\begin{table}[htb]
\caption{Production and properties of radioactive beams.}
\begin{ruledtabular}
\begin{tabular}{cccccc}
\multicolumn{2}{c}{Primary beam} & \multicolumn{4}{c}{Secondary
beam}  \\ \cline{1-2} \cline{3-6}
Ion  & Intensity & Ion & $E$ & Intensity & Purity \\
 & (pnA) &  & (MeV) & (pps) & (\%)  \\ \hline
$^{40}$Ar & 25 & $^{38}$S & 1547.5 & $ 2 \cdot 10^5$ & $>99$ \\
$^{48}$Ca & 15 & $^{40}$S & 1582.5 & $ 2 \cdot 10^4$ & $>95$ \\
\end{tabular}
\end{ruledtabular}
\protect \label{tab:beams}
\end{table}

\subsection{Target design and test}

Prior to the transient-field measurement a test run was performed to
ensure that the fast fragment beams at $\sim 40$ MeV/nucleon would
be slowed to the appropriate velocity regime in the target layers
and emerge from the target for downstream detection without severe
energy straggling. A 1540 MeV $^{38}$S beam was used for this
purpose.

Three targets were mounted, in turn, on a rotating target ladder and
the energy distribution of the emerging ions was measured with a
980~$\mu$m PIN detector placed 41 cm downstream of the target.
The targets were: (i) the 355 mg/cm$^2$ Au foil alone, (ii) the Au
(355 mg/cm$^2$) + Fe (100 mg/cm$^2$), and (iii) the Au (355
mg/cm$^2$) + Fe (110 mg/cm$^2$) target, which was subsequently
chosen for the $g$-factor measurement. The angle between the target
and the beam was varied to effectively increase the target
thickness. The transmission of the beam through the target and the
energy distribution of the transmitted ions were measured using the
PIN detector.
%
%
%
These measurements indicated that the Ziegler (1985) stopping powers
\cite{Zie85} for sulfur in Au and Fe targets were accurate to about
$\pm 5\%$, although they tend to underestimate the stopping in the
Au layer of the target and overestimate the stopping in the Fe
layer.

\begin{figure}
\resizebox{80mm}{!}{\includegraphics{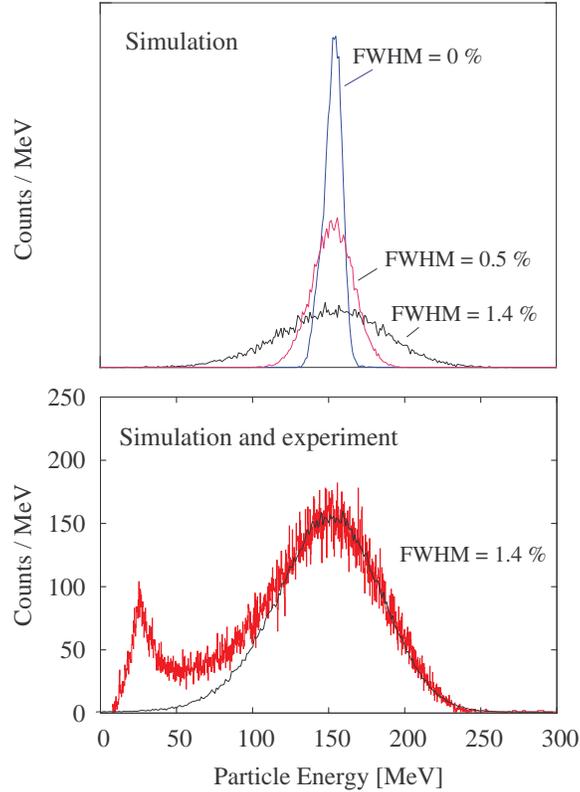}}
 \caption{(Color online) Spectrum of 1540 MeV $^{38}$S beam particles after passing
through 355 mg/cm$^2$ of Au followed by 100 mg/cm$^2$ of Fe. The
upper panel shows simulations that examine the effect of the
energy-width of the $^{38}$S beam, which is assumed to have a
Gaussian distribution about 1540 MeV with a specified Full Width at
Half Maximum (FWHM). These simulated spectra all have the same
number of events. The experimental spectrum in the lower panel
agrees with the simulations which assume the beam energy has a
Gaussian distribution with a FWHM of 1.4\% (i.e. a standard
deviation of 0.5\%). The equivalent momentum width is $\Delta p / p
= 0.7\%$.}\label{fig:testrun}
\end{figure}

The effect of the energy-width of the radioactive beam on the
spectrum of the emergent ions is examined in Fig.~\ref{fig:testrun},
which compares the particle spectrum measured in the test run for
the Au + Fe (100 mg/cm$^2$) target with Monte Carlo simulations made
with the code GKINT\_MCDBLS \cite{McDBLS}. In the simulations the
energy spread of the beam was treated as a Gaussian distribution
with its width specified by the Full Width at Half Maximum (FWHM).
In the case shown in Fig.~\ref{fig:testrun} the sulfur fragments
emerged with energies in the range from $\sim 80$~MeV to $\sim
200$~MeV. A similar post-target energy distribution was observed in
the $g$~factor measurements where both the beam energy and the
thickness of the Fe layer were increased slightly. It is evident
from Fig.~\ref{fig:testrun} that most of this energy-spread stems
from the energy width of the radioactive beam, which is determined
largely by the momentum acceptance, $\Delta p / p =0.5\%$ (FWHM),
set at the dispersive image slits of the A1900 spectrometer.

\subsection{Apparatus}

\begin{figure}[tb]
\resizebox{86mm}{!}{\includegraphics{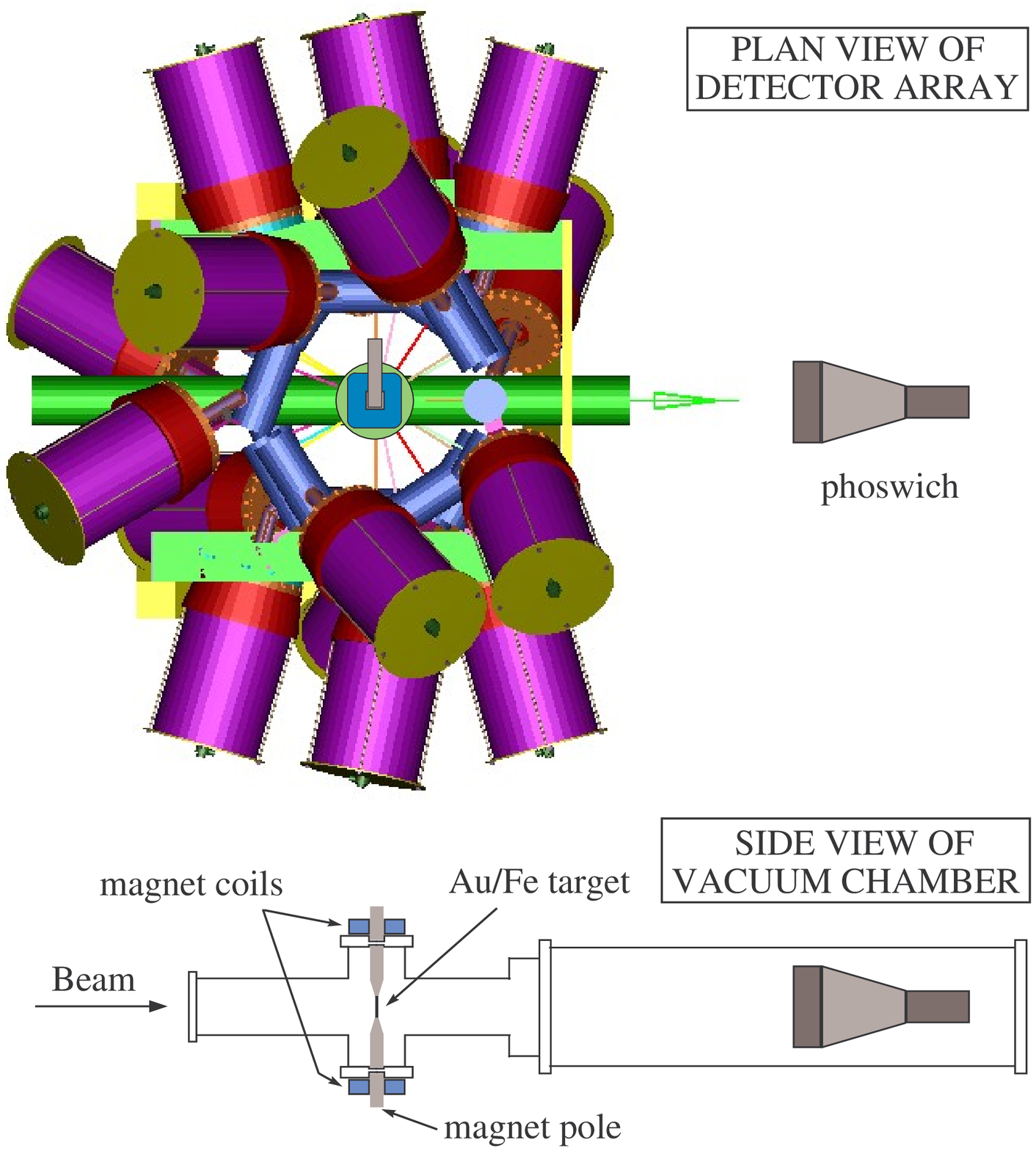}} \caption{(Color
online) Overhead view of the SeGA detectors and side view of the
target chamber showing the magnet, target and phoswich detector.
Projectiles were detected in the downstream phoswich detector.}
\label{fig:sideview2}
\end{figure}

Figure \ref{fig:sideview2} shows the experimental arrangement. The
radioactive beams were delivered onto the Au + Fe target, which had
dimensions $30 \times 30$ mm$^2$. The target was held between the
pole tips of a compact electromagnet that provided a magnetic field
of 0.11 T, sufficient to fully magnetize the Fe layer. As the iron
layer of the target is much thicker than those typically used in
transient-field $g$-factor measurements, the magnetization of a
piece was measured using the Rutgers Magnetometer \cite{Piq89}.
Fields of 0.062~T were sufficient to ensure saturation. To minimize
possible systematic errors, the external magnetic field was
automatically reversed every 600 s.

Table~\ref{paramtable} summarizes the properties of the 2$^+_1$
states and the key aspects of the energy loss of the sulfur beams in
the target, applicable for the $g$~factor measurements. The high-$Z$
Au target layer serves to enhance the Coulomb excitation yield and
slow the projectiles to under 800 MeV, while the thick iron layer
results in a long interaction time with the transient field,
maximizing the spin precession. The energies $\langle E_i \rangle$,
with which the sulfur ions enter the iron layer, were calculated
taking into account the energy loss measurements for the Au layer
alone, while the energies $\langle E_e \rangle$, with which the ions
emerge from the iron layer into vacuum, were determined from the
Doppler shifts observed in the $g$-factor measurements.

\begin{table*}
\caption{Nuclear parameters and reaction kinematics.
$B(E2)\!\!\uparrow \ = B(E2; 0^+_{\rm gs} \rightarrow 2^+_1)$.
$\langle E_{i,e} \rangle$ and $\langle v_{i,e} \rangle$ are the ion
kinetic energies and velocities at the entrance and exit of the iron
layer. The effective transient-field interaction time $t_{\rm eff}$
is evaluated for ions that decay after leaving the target. }
\begin{ruledtabular}
\begin{tabular}{*{9}c}
Isotope & E(2$^+_1$) & $B(E2)\uparrow$ & $\tau$(2$^+_1$) & $\langle
E_i \rangle$ & $\langle E_e \rangle$ & $\langle v_i/Zv_0 \rangle$ &
$\langle v_e/Zv_0 \rangle$ & $t_{\rm eff}$ \\
 & (keV) & (e$^2$fm$^4$) & (ps) & (MeV) & (MeV) & & & (ps)\\
\hline
$^{38}$S & 1292 & 235(30) & 4.9 & 762 & 123 & 1.75 & 0.71 & 2.98  \\
$^{40}$S & 904 & 334(36) & 21 & 782 & 145 & 1.73 & 0.75 & 2.99 \\
\end{tabular}
\end{ruledtabular}
\protect \label{paramtable}
\end{table*}

Projectiles scattered forward out of the target were detected with a
15.24 cm diameter plastic scintillator phoswich detector placed 79.2
cm downstream of the target position. The phoswich detector
consisted of a 750 $\mu$m layer of fast BC-400 scintillator and a
5.08 cm layer of slow BC-444 scintillator. The maximum scattering
angle, 5.5$^{\circ}$, limits the distance of closest approach to
near the nuclear interaction radius in both the Au and Fe target
layers. Positioning the particle detector downstream also lowers the
exposure of the $\gamma$-ray detectors to the radioactive decay of
the projectiles.

While the sulfur fragments do not penetrate beyond the fast
scintillator, the slow scintillator helps discriminate against more
penetrating radiation such as light ions produced in the secondary
target, or accompanying the beam, and $\beta$ decays in the phoswich
from the decay of the implanted radioactive beam. Along with the
fast-slow particle identification information from the phoswich, the
particle time-of-flight was also recorded with respect to the
cyclotron RF. Triggers from the radioactive decay of the beam
particles, which are much lower in energy than the beam particles,
were minimized by raising the threshold of the phoswich
discriminator.

For the $^{38}$S run, a circular Pb mask was placed on the phoswich
detector to block particles in the range $0^\circ \leq \theta \leq
2.5^{\circ}$ and hence lower the count rate by excluding those
scattering angles where the Rutherford cross section is large but
the Coulomb excitation cross section is small. The mask helped
reduce random particle-$\gamma$ coincidences and pileup events in
the phoswich detector. No mask was used for the $^{40}$S run since
the particle rate was low enough for pileup to be negligible.

\begin{figure}
\resizebox{86mm}{!}{\includegraphics{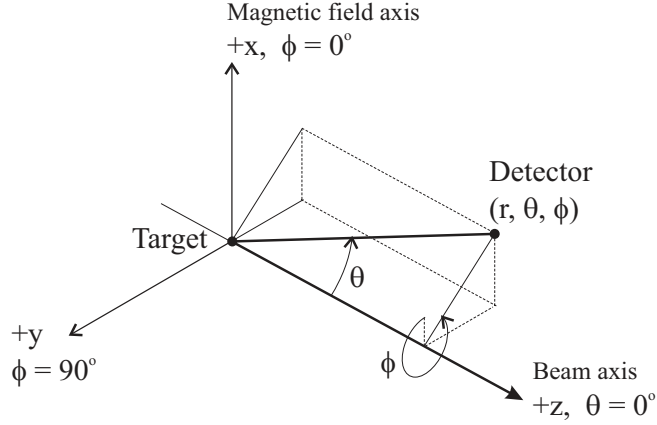}} \caption{Relation
between beam axis, magnetic field axis, and the coordinate systems
used in the present experiment and its analysis. }\label{fig:angles}
\end{figure}

To detect de-excitation $\gamma$ rays, the target chamber was
surrounded by 14 HPGe detectors of the Segmented Germanium Array
(SeGA) \cite{Muel2001}. The SeGA detectors were positioned with the
crystal centers 24.5 cm from the target position. Six pairs of
detectors were fixed at symmetric angles $(\pm\theta,\phi) =
(29^{\circ}, 90^\circ)$, $(40^{\circ}, 131^\circ)$, $(60^{\circ},
61^\circ)$, $(139^{\circ}, 46^\circ)$, $(147^{\circ}, 143^\circ)$,
and $(151^{\circ}, 90^\circ)$, where $\theta$ is the polar angle
with respect to the beam axis and $\phi$ is the azimuthal angle
measured from the vertical direction, which coincides with the
magnetic field axis. Figure \ref{fig:angles} indicates the
coordinate frame and definitions of the angles. To make a connection
with the notation used for conventional transient-field measurements
\cite{Kol1980,Spe2002}, the locations of the pair of detectors at
the spherical polar angles $(\theta,\phi)$ and
$(\theta,\phi+180^\circ)$ are written as $(\pm \theta, \phi)$; thus
if $(+\theta,\phi) = (|\theta|,\phi)$, then $(-\theta,\phi) =
(|\theta|,\phi + 180^\circ)$. Each $\pm \theta$ pair is in a plane
that passes through the center of the target.

Two additional detectors were placed at $(\theta, \phi) =
(90^{\circ}, 112^\circ)$ and $(24^{\circ}, 0^\circ)$ to assist the
measurement of the angular correlation. All 14 detectors were used
to measure the $\gamma$-ray angular correlations concurrently with
the precessions. Since the precession angles are small, the
unperturbed angular correlation can be reconstructed by adding the
data for the two directions of the applied magnetic field.

The positions of the SeGA detectors with respect to the target
position were measured to an accuracy of 2 mm using a theodolite
system.  This information was used to find the actual crystal
locations in conjunction with the SeGA crystal segment positions
measured by Miller {\em et al.} \cite{Mill2002}. The 32-fold
segmentation of the detectors improves the position determination of
the $\gamma$ ray from the entire crystal length (8 cm) to near the
segment length (1 cm), which is needed for Doppler corrections due
to the high velocity of the emitting nuclei. The angle of
$\gamma$-ray emission was deduced from the position of the detector
segment that registered the highest energy deposition. With this
algorithm the position resolution is near (but does not reach) the
segment length.

The master trigger for the data acquisition was set to record
particle-$\gamma$ coincidences as well as down-scaled particle
singles. Thirty three energy signals were recorded for each
$\gamma$-ray detector, corresponding to the 32 segments and the
central contact. To differentiate between the signals from the fast
and slow scintillator, the pulse from the phoswich detector was
charge-integrated over the whole signal and over the tail (slow)
part of the signal in separate QDC channels. Time differences were
recorded between the phoswich detector and the $\gamma$-ray
detectors. To assist with particle identification, the
time-of-flight spectrum was also recorded for particles striking the
phoswich detector. Finally, each event included a tag that
identified the direction of the external magnetic field.

\section{Experimental results and analysis} \label{sect:results}

\subsection{Particle and $\gamma$-ray spectra} \label{subsect:spectra}

\begin{figure}
\resizebox{86mm}{!}{\includegraphics{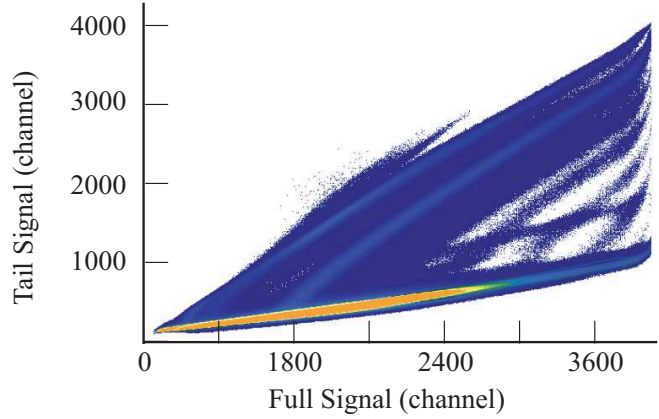}} \caption{(Color
online) Phoswich particle identification plot recorded during the
$^{38}$S run. Pulse-shape discrimination was obtained by plotting
the integrated charge for the slow (tail) signal versus the full
charge-integrated signal. }\label{fig:phoswich}
\end{figure}

A particle identification plot obtained with the phoswich detector
during the $^{38}$S measurement is shown in Fig.~\ref{fig:phoswich}.
This spectrum was obtained by plotting the integrated charge for the
slow (or tail) part of the phoswich signal versus the integrated
charge for the whole signal. Most of the intensity corresponds to
the sulfur projectiles. It has a small tail component and therefore
occurs in a band located along the full signal axis. The upsloping
lines of intensity that deviate from this heavy-ion band are due to
low-$Z$ particles that punch through into the slow scintillator.
Such ions accompany the beam, probably originating at the acrylic
Image-2 wedge in the A1900 spectrometer; they are not predominantly
produced in the secondary target.

\begin{figure}
\resizebox{86mm}{!}{\includegraphics{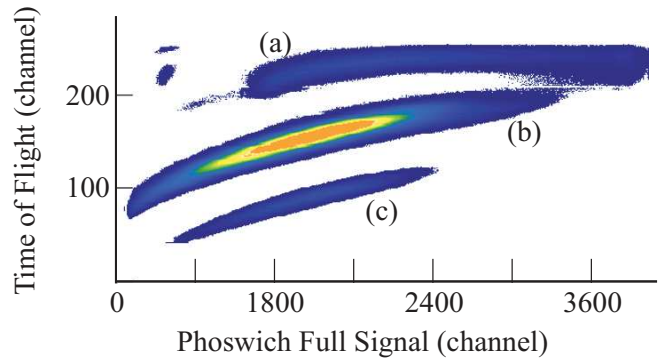}} \caption{(Color
online) Time-of-flight (between the cyclotron RF and the phoswich
detector) versus the  phoswich energy recorded during the $^{38}$S
run. The labeled features are (a) light particles registered in the
slow scintillator, (b) $^{38}$S ions in coincidence with $\gamma$
rays, and (c) down-scaled events. See text for further details.
}\label{fig:phoswichRF}
\end{figure}

The spectrum in Fig.~\ref{fig:phoswichRF} shows the time-of-flight
between the cyclotron RF and the phoswich trigger pulse versus the
phoswich energy signal (integration of the full pulse shape). Three
distinct features are evident in this spectrum. The upper feature,
labeled (a), corresponds to events which cause a large amount of
slow scintillation in the phoswich; they are the light particles
discussed already. The middle region, labeled (b), corresponds to
phoswich events triggered by sulfur projectiles that are in
coincidence with $\gamma$ rays, and the lower region, labeled (c),
is made up of downscaled phoswich events. (The downscaler module
introduces an additional delay that shifts this group of events away
from the phoswich-$\gamma$ coincidences.)

\begin{figure}[tbp]
\begin{center}
\begin{tabular}{cc}
\resizebox{44mm}{!}{\includegraphics{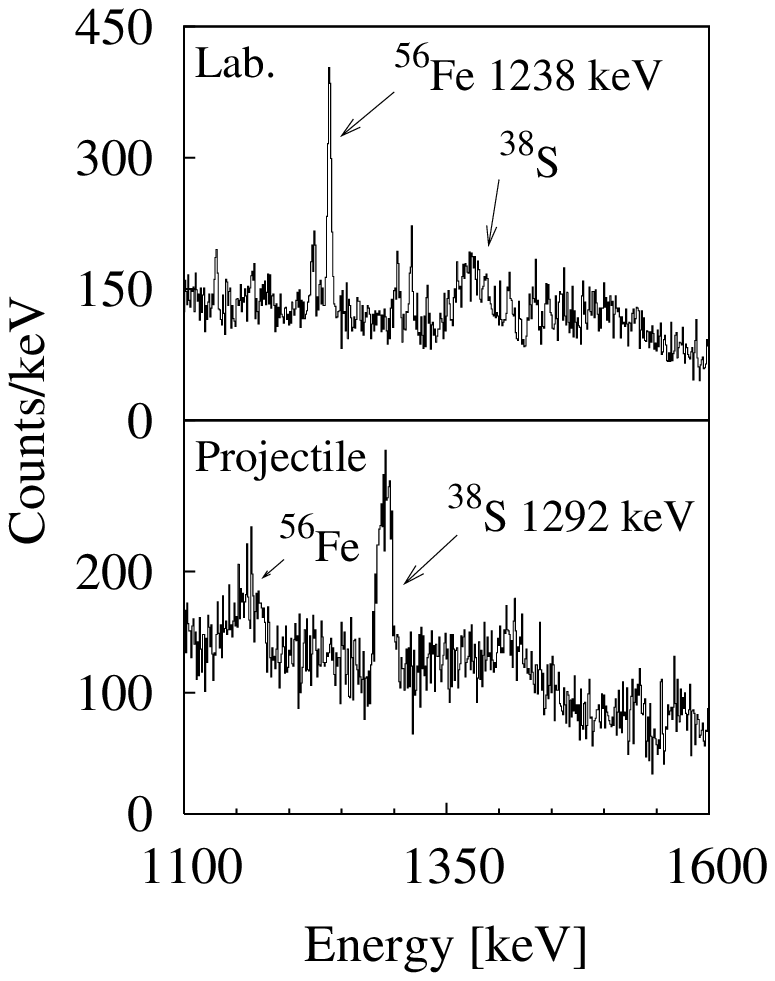}} &
\resizebox{40mm}{!}{\includegraphics{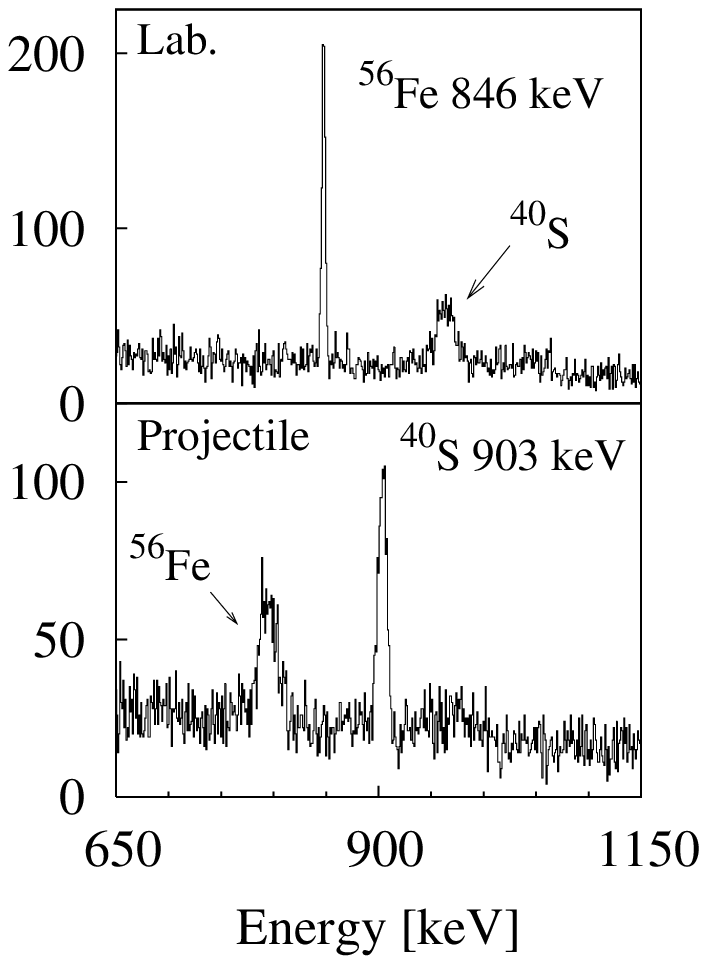}}
\end{tabular}
\caption{Random-subtracted lab frame (top) and Doppler-corrected
(bottom) $\gamma$-ray spectra at $\theta=40^{\circ}$ for $^{38}$S
(left) and $^{40}$S (right). The sulfur and iron peaks are labeled.
The broad feature to the right of the 1292 keV line is mainly its
Doppler tail, due to decays within the target, although there is a
contribution from a contaminant line as well. The line at 1238 keV
is the $4 \rightarrow 2$ transition emitted by $^{56}$Fe nuclei at
rest.} \label{fig:3840Sspectra}
\end{center}
\end{figure}

Gamma-ray spectra gated on sulfur recoils were produced and
corrected for random coincidences by subtracting spectra gated on
the appropriate regions of the particle-$\gamma$ time spectra. Both
the particle identification and time-of-flight information from the
phoswich were used to select the events of interest. Spectra were
also created for each $\gamma$-ray detector and for each direction
of the magnetic field (up/down). Examples of the random-subtracted
spectra for the lab-frame are given in the upper panels of
Fig.~\ref{fig:3840Sspectra}. From the measured Doppler shift of the
deexcitation $\gamma$ rays in the laboratory frame, the average
after-target ion velocities were determined to be 0.083~c for
$^{38}$S and 0.088~c for $^{40}$S, (i.e. $v/Zv_0 = $ 0.71 and 0.75,
respectively). The velocity distribution of the exiting $^{40}$S
ions was also measured by shifting the phoswich detector by $\pm
15$~cm from its normal position and observing the change in the
flight times of the projectiles. These procedures firmly establish
that the sulfur ions were slowed through the peak of the TF strength
at $Zv_0$ into the region where it has been well characterized
\cite{Spe2002,Stuc2004a}.

Doppler-corrected spectra were also produced using the angular
information from the SeGA detector segments and the particle energy
information from the phoswich detector on an event-by-event basis,
which is essential because of the spread in particle velocities. The
lower panels of Fig.~\ref{fig:3840Sspectra} show examples of the
Doppler-corrected $\gamma$-ray spectra. The peak to background ratio
seems to be better for the $^{40}$S measurement because (i) there
are more counts in the $^{38}$S measurement, which gives an overall
higher count baseline, and (ii) the combination of the higher
$\gamma$-ray energy and shorter level lifetime in $^{38}$S result in
a broader Doppler-corrected peak.

\begin{figure}[tbp]
\resizebox{86mm}{!}{\includegraphics{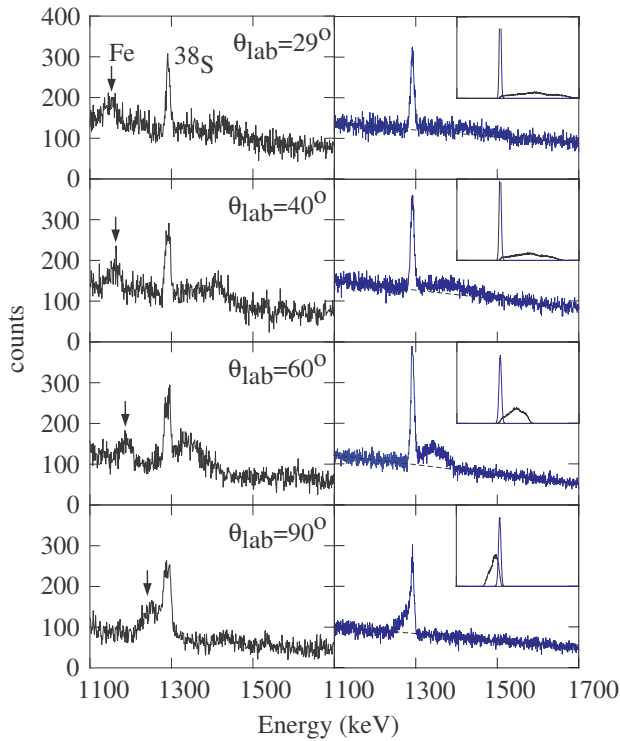}}
 \caption{(Color online) Random-subtracted Doppler-corrected $\gamma$-ray spectra from the
$^{38}$S measurement are shown on the left. The 1238 keV line
emitted by $^{56}$Fe nucleus at rest in the lab frame are shifted
towards the $^{38}$S peak by the Doppler correction as the detection
angle approaches 90$^\circ$. The right panels show simulated spectra
used to evaluate the impact of the Doppler Broadened Line Shape on
the angular correlation and precession measurements. The insets show
the energy range from 1150 to 1550 keV. These simulations, which do
not include background, indicate the relative contributions of
decays in the target (broad peak) and decays in vacuum (sharp peak),
which overlap. The experimental peak is broader than that in the
simulations because a simplified model is used for the $\gamma$-ray
detectors. } \label{fig:dbls}
\end{figure}

Most of the excited nuclei decay in vacuum after leaving the target
material, however for $^{38}$S a significant number also decay at
higher velocities, extending up to the secondary beam velocity,
whilst still within the target. The long Doppler tail becomes almost
indistinguishable from background at extreme forward and backward
angles, but is clear at $\theta = 40^\circ$, $60^\circ$ and
$90^\circ$. It was established through Doppler Broadened Line Shape
(DBLS) calculations based on a Monte Carlo approach that the angular
correlations and nuclear precessions can be determined accurately by
an analysis of the vacuum-flight peak alone, without the need to
include the Doppler tail. Examples of the DBLS calculations and
comparisons with the experimental spectra are shown in
Fig.~\ref{fig:dbls}. From these spectra it can be seen that the
proportion of the $\gamma$-ray peak that corresponds to decays
within the target is very small, and only weakly dependent on the
detection angle. Note that the present calculations of the
lineshapes do not account for the possibility that a $\gamma$-ray
event may be assigned to the wrong segment of the detector. Having
established that the correct angular correlation and $g$-factor
results are obtained by analyzing only the $\gamma$-ray peak, and
excluding the tail, no attempt was made to obtain a quantitative fit
to the observed $\gamma$-ray line shapes. Further details of these
calculations will be given below and elsewhere \cite{McDBLS}.

The 2$^+$ peak areas averaged 925 counts/detector per field
direction for $^{38}$S and 400 counts/detector per field direction
for $^{40}$S, in each of the six angle pairs of SeGA detectors used
for extracting the precessions.

\subsection{Angular correlations and precessions}


In the rest frame of the nucleus, the perturbed angular correlation
is given by
\begin{eqnarray}
W(\theta, \phi, \pm \Delta \Theta) &=& 4 \pi
\sum_{k,q} \frac{a_k Q_k G_k}{(2k+1)} (-1)^q  \nonumber \\
 & \times & Y^k_{q}(|\Delta
\Theta|,\pm \pi/2) Y^k_{- q}(\theta,\phi) \label{eq:pac}
\end{eqnarray}
where $(\theta,\phi)$ are the spherical polar angles at which the
$\gamma$-ray is detected and $\pm \Delta \Theta$ is the
transient-field precession angle for the two directions of the
magnetic field; $k = 0,2,4$ and $q$ takes integer values such that
$-k \leq q \leq k$. As shown in Fig.~\ref{fig:angles}, the beam
direction defines $\theta = 0$ and the magnetic-field direction
defines $\phi = 0$. The unperturbed angular correlation has symmetry
about the beam axis and reduces to the $\phi$-independent form:
\begin{equation}
W(\theta) = 1 + \sum_{k = 2,4} a_k Q_k G_k P_k(\cos \theta).
\label{AC}
\end{equation}

The $a_k$ coefficients, which depend on the particle detector
geometry, the spins of the initial and final nuclear states and the
multipolarity of the $\gamma$-ray transition, can be evaluated from
the theory of Coulomb excitation \cite{Bert2003}. The correction
factors for the finite solid angles of the $\gamma$-ray detectors,
$Q_k$, are near unity in the present work. The deorientation
coefficients, $G_k$, account for the effect of hyperfine fields
experienced by ions that recoil into vacuum carrying atomic
electrons.

These recoil in vacuum effects were evaluated based on measured
charge-state fractions for sulfur ions emerging from iron foils with
energies between 92 and 236 MeV \cite{Stuc2005c}. The large free-ion
hyperfine interactions of hydrogen- and lithium-like ions
\cite{Gold1982}, which quickly reach hard-core values for nuclear
lifetimes of a picosecond or more, are dominant. To a good
approximation, the deorientation coefficients can be expressed as
$G_k = 1 - (Q_H + Q_{Li})b_k$, where the fraction of ions with one
and three electrons is $Q_H + Q_{Li}$ and $b_k = k(k+1)/(2I + 1)^2$
for nuclear spin $I$ and $k=2,4$.  Using the measured charge-state
fractions gives $G_2 = 0.90$ and $G_4 = 0.65$.

The angular correlations were calculated with the program GKINT
\cite{Stuc2005b}.
%
%
For each layer of the target, the code GKINT performs integrals over
the solid angle of the particle detector and over the energy loss of
the beam in the target. The average $a_k$ coefficients, for example,
are evaluated as
\begin{eqnarray}\protect\label{eq:ak}
\langle a_k \rangle &=& \frac{2 \pi}{\langle \sigma \cdot t_{\rm
tgt} \rangle} \int_{0}^{t_{\rm tgt}} \int_{\theta_{p}^{\rm
min}}^{\theta_{p}^{\rm max}} a_k(\theta_p, E[z])  \nonumber \\
 & \times& \frac{d \sigma }{d \Omega }(\theta_p , E[z])  \; \sin \theta_p d
\theta_p d z,
\end{eqnarray}
where
\begin{equation}
\langle \sigma \cdot t_{\rm tgt} \rangle = 2 \pi \int_{0}^{t_{\rm
tgt}} \int_{\theta_{p}^{\rm min}}^{\theta_{p}^{\rm max}} \frac{d
\sigma }{d \Omega }(\theta_p , E[z])  \; \sin \theta_p d \theta_p d
z,
\end{equation}
where $\frac{d \sigma }{d \Omega }(\theta_p , E[z])$ denotes the
Coulomb excitation cross section as a function of the scattering
angle $\theta_p$ and the projectile energy $E[z]$, which varies with
the depth $z$ through the target layer. The integrals are evaluated
numerically using Simpson's rule. To evaluate other average
quantities of interest such as the average energy of excitation, or
the average transient-field precession, the quantity to be averaged
replaces $a_k$ in an expression of the same form as
Eq.~(\ref{eq:ak}).

\begin{figure}[btp]
\resizebox{86mm}{!}{\includegraphics{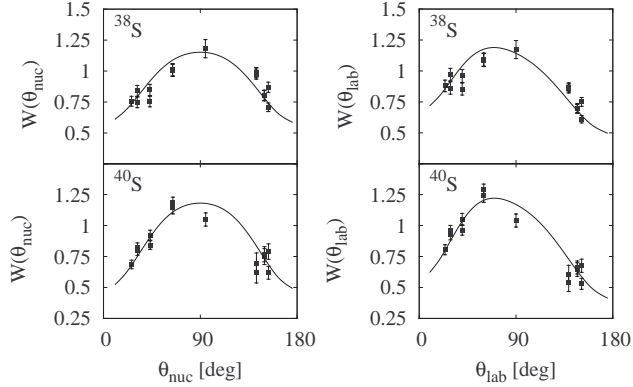}} \caption{Angular
correlations for $^{38}$S and $^{40}$S. The left panels show the
angular correlation in the frame of the projectile nucleus, while
the right panels show the same angular correlations in the lab
frame. The Lorentz boost causes a shift in the effective detection
angle and an angle-dependent change in the solid angle. Data are
normalized to the calculated angular correlation. The difference in
anisotropy stems from the alignment produced by Coulomb excitation,
which depends on the ratio of $E(2^+)$ to the beam velocity (i.e.
the adiabacity parameter, $\xi$, see \protect \cite{Bert2003} and
references therein.) } \label{3840Sangdist}
\end{figure}

Comparisons between the experimental and theoretical angular
correlations are made in Fig~\ref{3840Sangdist}. To indicate the
effect of the Lorentz transformation (see e.g.
\cite{Stu2003,Oll2003}) the angular correlations are shown in both
the laboratory frame and the projectile frame. Good agreement was
found between the calculated $\gamma$-ray angular correlations and
the data.

\begin{figure}[tbp]
\resizebox{86mm}{!}{\includegraphics{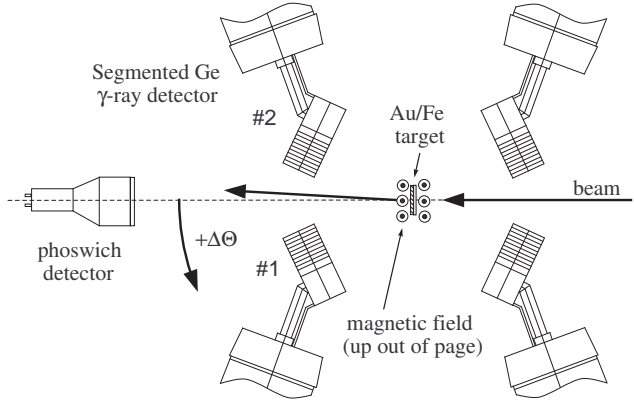}} \caption{Schematic
view of the experimental arrangement from above showing only the
four SeGA detectors perpendicular to the magnetic field axis. The
direction of positive precession angles, $\Delta \Theta$, is
indicated. } \label{fig:4detectors}
\end{figure}

To describe the procedure for the extraction of the nuclear
precession angle $\Delta \Theta$ it is helpful to begin with the 4
detectors in the plane through the target that is perpendicular to
the magnetic field direction, as illustrated in
Fig.~\ref{fig:4detectors}. For these detectors $\Delta \Theta =
\Delta \theta$ and the data analysis is identical to that in a
conventional transient-field $g$-factor measurement
\cite{Kol1980,Spe2002,Jak1999,Man2001}. The magnetic field causes a
rotation of the angular correlation pattern around the $x$-axis,
with positive precession angles for a right-handed co-ordinate frame
in the direction indicated. Thus to first order in $\Delta \Theta$
we have
\begin{equation}
W\uparrow \downarrow = W(\theta \mp \Delta \Theta) \simeq W(\theta)
\mp \Delta \Theta \frac{dW}{d \theta} ,
\end{equation}
where the arrows indicate the direction of the external magnetic
field. The precession angle is then determined from the double ratio
of counts in a pair of detectors for each field direction.
Considering, for example, the pair of detectors labeled 1 and 2 in
Fig.~\ref{fig:4detectors}, the relevant experimental double ratio
$\rho$ is
\begin{equation}
\rho = \sqrt{\frac{N_1 \uparrow}{N_1 \downarrow} \cdot \frac{N_2
\downarrow}{N_2 \uparrow}}.
\end{equation}
Since the detector efficiencies and running times cancel out, $\rho$
and $\Delta \Theta$ are related by
\begin{equation}
\rho = \frac{1 - S(\theta) \Delta \Theta}{1 + S(\theta) \Delta
\Theta},
\end{equation}
where
\begin{equation}
S(\theta) = \left . \frac{1}{W} \frac{dW}{d \theta} \right |_\theta.
\end{equation}
It is conventional to define the `effect',
\begin{equation}
\epsilon = (1 - \rho)/(1 + \rho),
\end{equation}
so that
\begin{equation}
\Delta \Theta = \epsilon / S.
\end{equation}
These expressions apply only for the 4 detectors in the $\phi =
90^\circ$ (horizontal) plane. We now generalize to include the
detector pairs that are not in this plane. By expanding
Eq.~(\ref{eq:pac}) to first order in $\Delta \Theta$ it can be shown
that
\begin{equation}
W\uparrow \downarrow = W(\theta, \phi, \mp \Delta \Theta) \simeq
W(\theta) \mp \Delta \Theta \sin \phi \frac{dW}{d \theta} .
\end{equation}
It follows that data analysis for a pair of detectors placed at the
spherical polar angles $(\theta,\phi)$ and $(\theta,\phi+180^\circ)$
can proceed exactly as in the familiar case, but with the definition
\begin{equation}
S(\theta, \phi) = S(\theta) \sin \phi.
\end{equation}
Finally, it remains to note that the effect of the Lorentz
transformation must be taken into account by evaluating $S$ at the
appropriate angle in the rest frame of the nucleus that corresponds
to the laboratory detection angle \cite{Stuc2005b}.

The precession results are summarized in
Table~\ref{tab:precessions}. The two pairs of detectors at $\theta =
29^{\circ}$ and 151$^{\circ}$, which were located in the horizontal
plane perpendicular to the external magnetic field direction, near
the angle of maximum slope of the $\gamma$-ray angular correlation,
are most sensitive to the nuclear precession.

To cover the uncertainties in the recoil in vacuum corrections and
the possibility that some nuclear interference might affect the
Coulomb excitation process for the (small) fraction of collisions
that approach the nuclear interaction radius, an uncertainty of $\pm
10\%$ has been assigned to the $S$ values.

\begin{table*}
\caption{Precession results. }
\begin{ruledtabular}
\begin{tabular}{*{10}c}
\multicolumn{4}{c}{detectors} & \multicolumn{3}{c}{$^{38}$S} &
\multicolumn{3}{c}{$^{40}$S}\\ \cline{1-4}\cline{5-7} \cline{8-10}
Pair & $\theta ^\circ$ & $\phi_1^\circ$ & $\phi_2^\circ$ &
$\epsilon$ $(\times 10^3)$ & $S$ & $\Delta
\Theta$ (mrad)&  $\epsilon$ $(\times 10^3)$ & $S$ & $\Delta \Theta$ (mrad)\\
\hline
1 & $29$  & 270 & 90  & --50(20) & +0.79(8) & --64(27) & --17(35) & +0.86(9) & --20(41)\\
2 & $151$ & 270 & 90  & +26(23) & --0.80(8) & --33(29) & --4(35)  & --0.88(9) & +5(40)\\
3 & $139$ & 226 & 46  & --10(23) & --0.54(5) & +19(43) & +7(28)  & --0.58(6) & --12(48)\\
4 & $147$ & 323 & 143 & --7(27)  & --0.48(5) & +15(57) & --24(31) & --0.52(5) & +46(60)\\
5 & $40$  & 311 & 131 & --41(19) & +0.50(5) & --83(39) & +26(32) & +0.53(5) & +49(60)\\
6 & $60$  & 241 & 61  & --18(15) & +0.26(3) & --69(58) & +1(28)  &
+0.27(3) & +4(103)\\ \cline{7-7} \cline{10-10}
\multicolumn{6}{l}{average:} &$-43(15)$ & & &$+5(21)$ \\
\end{tabular}
\end{ruledtabular}
\protect \label{tab:precessions}
\end{table*}

\subsection{Transient-field calibration and $g$~factor results}

\begin{figure}
\resizebox{86mm}{!}{\includegraphics{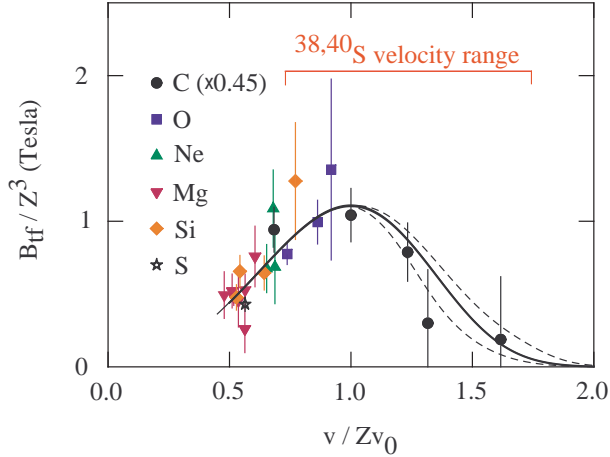}} \caption{(Color
online) Transient field parametrization (solid line) for
high-velocity light ions, from Ref.~\cite{Stuc2004a}. Data points
are measured transient-field strengths reported in the literature.
See \protect ~\cite{Spe2002,Stuc2004a,Stuc2005a} for further details
and references.}
\protect \label{fig:TFparam}
\end{figure}

An evaluation of
\begin{equation}
\Delta \Theta / g = (-\mu_{\rm N}/\hbar) \int B_{\rm tf} dt
\end{equation}
is required to extract the experimental $g$~factors. For light ions
($Z \leq 16$) traversing iron and gadolinium hosts at high velocity,
the dependence of the TF strength on the ion velocity, $v$, and
atomic number, $Z$, can be parametrized \cite{Stuc2004a,Stuc2005a}
as
\begin{equation}
B_{\rm tf}(v,Z) = A Z^P (v/Zv_0)^2 {\rm e}^{-
\frac{1}{2}(v/Zv_0)^4}, \protect \label{eq:aes-param}
\end{equation}
where $v_0 = c/137$ is the Bohr velocity. This is a model-based
parametrization that takes account of the known physics of the
transient field (see \cite{Spe2002,Stuc2004a,Stuc2005a} and
references therein). A fit to data for iron hosts yielded
$A=1.82(5)$~T with $P=3$ \cite{Stuc2004a}. The experimental data and
the adopted parametrization are shown in Fig.~\ref{fig:TFparam}. The
velocity range sampled in the present experiments is also indicated.
Although the data are sparse in the high velocity region, there can
be no dispute about the general trend, and that the maximum TF
strength is reached when the ion velocity matches the $K$-shell
electron velocity, $v = Zv_0$. Also shown in Fig.~\ref{fig:TFparam}
are two alternative parametrizations of the field strength in the
region $v > Zv_0$ chosen to give an indication of the uncertainty in
the transient-field calibration. Compared with the adopted
parametrization, these extrapolations give values of $\Delta
\Theta/g$ that differ by $\sim \pm 12\%$.

Calculations of $\Delta \Theta/g$ were performed using the code
GKINT to take into account the incoming and exiting ion velocities,
the energy- and angle-dependent Coulomb excitation cross sections in
both target layers, the excited-state lifetimes, and the
parametrization of the TF strength in Eq.~(\ref{eq:aes-param}). The
results and the $g$~factors extracted are given in
Table~\ref{tab:results}.

The $g$~factor results are  not very sensitive to the somewhat
uncertain behavior of the transient field at the highest velocities
because (i) the ions spend least time interacting with the TF at
high velocity and (ii) the TF strength near $2Zv_0$ is very small.
Furthermore, the positive $g$~factor in $^{38}$S and the essentially
null effect for $^{40}$S are both firm observations, independent of
the transient-field strength. The $g$~factor of $^{38}$S is almost 3
standard deviations from zero.



The experimental uncertainties assigned to the $g$~factors are
dominated by the statistical errors in the $\gamma$-ray count
ratios, with small contributions from the angular correlation (10\%)
and transient field calibration (12\%) added in quadrature.

\begin{table}
\caption{Experimental results. The spin precession per unit
$g$~factor, $(\Delta \Theta/g)_{\rm calc}$, is evaluated for ions
that decay after leaving the target. The experimental $g$~factor is
given by $g = \Delta \Theta_{\rm exp} / (\Delta \Theta / g)_{\rm
calc}$. }
\begin{ruledtabular}
\begin{tabular}{*{4}c}
Isotope & $(\Delta\Theta/g)_{\rm calc}$ & $\Delta \Theta_{\rm exp}$ & $g$\\
 & (mrad) & (mrad)\\
\hline
$^{38}$S &  $-330(41)$ & $-43(15)$ & $+0.13(5)$ \\
$^{40}$S &  $-339(42)$ & $+5(21)$ & $-0.01(6)$\\
\end{tabular}
\end{ruledtabular}
\protect \label{tab:results}
\end{table}

\subsection{Monte Carlo simulations}

In the analysis of the data presented above an analytical formalism
has been used to evaluate the average angular correlation
coefficients, the transient-field precession per unit $g$~factor,
and other quantities of interest.

To obtain a more detailed insight into the experiment and data
analysis procedures it is helpful to use a Monte Carlo approach to
model the experiment. The code GKINT\_MCDBLS was written to generate
event data for particle-gamma coincidences that can be sorted and
correlated with calculated quantities such as the transient-field
precession. The effect of the energy spread on the incident beam on
the particle spectrum beyond the target was discussed above, and
examples of the calculated Doppler broadened $\gamma$-ray line
shapes were presented in Fig.~\ref{fig:dbls}. Here we examine the
behavior of the precession angle and the angular correlation
coefficients $a_2$ and $a_4$ as a function of the detected
$\gamma$-ray energy.

Aside from the statistical uncertainties, which can be minimized by
increasing the number of events in the simulation, the Monte Carlo
approach allows for a rigorous treatment of the decay-in-flight and
vacuum deorientation effects. It also properly includes the
contributions to the $\gamma$-ray peak that is analyzed to extract
the $g$~factor, stemming from excitation in the Au and Fe layers of
the target, and which cannot be separated experimentally.

\begin{figure}
\resizebox{86mm}{!}{\includegraphics{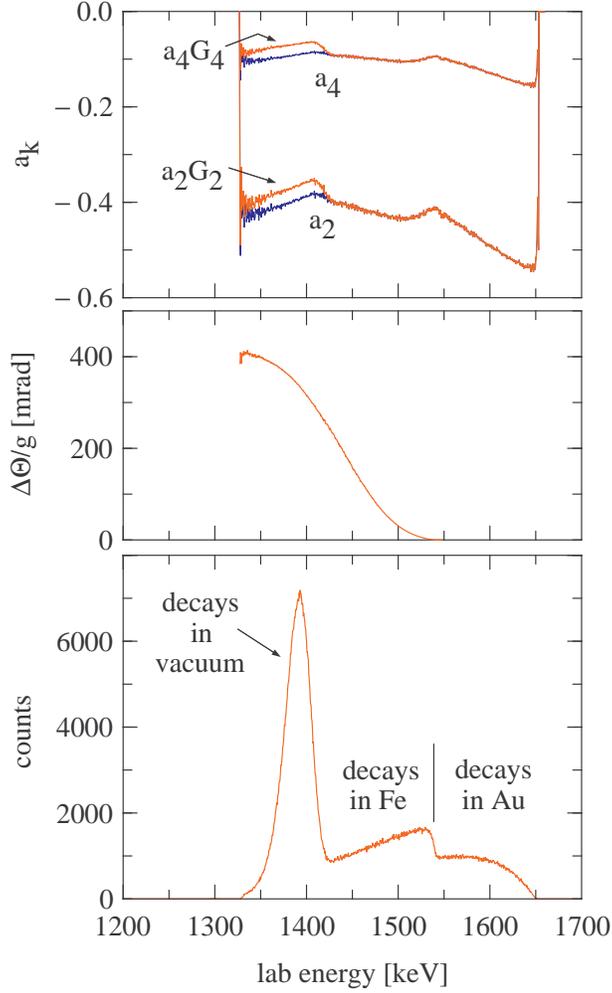}} \caption{(Color
online) Monte Carlo simulations (10$^6$ events). Lower panel:
Doppler broadened $\gamma$-ray line shape at 29$^\circ$ in the
laboratory frame. An equivalent spectrum in the projectile frame is
shown in the uppermost panel of Fig.~\protect \ref{fig:dbls}. Much
of the energy width in the laboratory frame stems from the spread in
the projectile velocity, which must be Doppler-corrected
event-by-event using the energy from the phoswich detector (section
\protect \ref{subsect:spectra}). Middle panel: The transient-field
precession per unit $g$~factor as a function of the emitted
$\gamma$-ray energy. Upper panel: Angular correlation coefficients
as a function of the emitted $\gamma$-ray energy. The effect of
vacuum deorientation is indicated by the (red) line labeled $a_k
G_k$.} \protect \label{fig:MCphi}
\end{figure}

Figure~\ref{fig:MCphi} shows how the angular correlation
coefficients, $a_2$ and $a_4$, and the transient field precession,
$\Delta \Theta/g$, vary as a function of the $\gamma$-ray line shape
in the laboratory frame. A detector at 29$^\circ$ in the laboratory
frame was chosen for this comparison. The bump in the line shape
near 1540 keV (lower panel) is due to the change in stopping powers
as the $^{38}$S ions pass into the Fe layer. As can be seen in the
middle panel, there is no transient-field precession in the highest
part of the Doppler tail (above the bump). This part of the line
shape corresponds to $^{38}$S ions that are excited and then decay
in the Au layer of the target, before reaching the iron layer. As
the sulfur ions slow and decay within the iron layer the
transient-field precession $\Delta \Theta/g$ increases, reaching its
maximum values for those ions that do not decay until they leave the
iron layer and emerge into vacuum. The upper part of
Fig.~\ref{fig:MCphi} shows the variation of both $a_k$ and $a_k
G_k$, where $G_k$ is the vacuum deorientation coefficient. Clearly,
the vacuum deorientation effect applies only to those ions that
decay in vacuum, and hence has an effect only in the peak part of
the line shape. Aside from the vacuum deorientation effect, the
variation in the magnitude of the $a_k$ coefficients as a function
of the Doppler shift has its origin in the alignment produced by
intermediate energy Coulomb excitation \cite{Bert2003,Stuc2005b}.
There is a pronounced dependence on the beam energy, such that the
alignment is reduced for those projectiles that are excited after
losing energy within the target.

At present the variation of the $\gamma$-ray intensity across the
acceptance of the detector is not included, and the response of the
segmented Ge detector is treated approximately. For example, it is
assumed that the hit is always assigned to the correct segment.
These approximations are not significant in the present context.

Because the $^{38}$S nuclei can be excited in either the Au or Fe
layers of the target, and many of them decay before they leave the
target and emerge into vacuum, the analytic (GKINT) calculations of
$a_2G_2$, $a_4G_4$ and $\Delta \Theta/g $ for the vacuum flight peak
required some simplifying approximations (mainly to estimate the
relative contributions of excitation in the two target layers). The
Monte Carlo (GKINT\_MCDBLS) calculations, however, enable a simple
and rigorous evaluation of these quantities of interest. The
disadvantage is that the Monte Carlo calculation is very time
consuming. It was found that the difference between the two
approaches was small in our case. As well as providing a deeper
insight into the experiment, the Monte Carlo calculations therefore
support our use of the approximate analytic calculations, which
introduce negligible error compared with the statistical
uncertainties in the experiment.

\section{Shell model and discussion} \label{sect:discussion}

\subsection{Shell model calculations}

Shell model calculations were performed for $^{36}_{16}$S$_{20}$,
$^{38}_{16}$S$_{22}$ and $^{40}_{16}$S$_{24}$, and their isotones
$^{38}_{18}$Ar$_{20}$, $^{40}_{18}$Ar$_{22}$ and
$^{42}_{18}$Ar$_{24}$, using the code OXBASH \cite{oxbash} and the
$sd$-$pf$ model space where (for $N\geq20$) valence protons are
restricted to the $sd$ shell and valence neutrons are restricted to
the $pf$ shell. The Hamiltonian was that developed in
Ref.~\cite{Numm2001} for neutron-rich nuclei around $N=28$, i.e. the
SDPF-NR interaction \cite{Caur2005}. These calculations reproduce
the energies of the low-excitation states to within 200 keV. With
standard effective charges of $e_p \sim 1.5$ and $e_n \sim 0.5$ they
also reproduce the measured $B(E2)$ values. The relevant results for
the 2$^+_1$ states are summarized in Table~\ref{tab:E2theory}, where
they are labeled SDPF. For the purposes of the following discussion,
the $B(E2)$ and $Q(2^+)$ values are also presented in terms of the
equivalent deformation parameter $\beta$. These electric quadrupole
properties have been calculated and discussed by Retamosa {\em et
al.} \cite{Reta1997}. Although the shell model Hamiltonian has been
improved since their work, the theoretical $B(E2)$ and deformation
values agree closely with the values in Table~\ref{tab:E2theory},
and their discussion remains relevant.

\begin{figure}[btp]
\begin{center}
\resizebox{75mm}{!}{\includegraphics{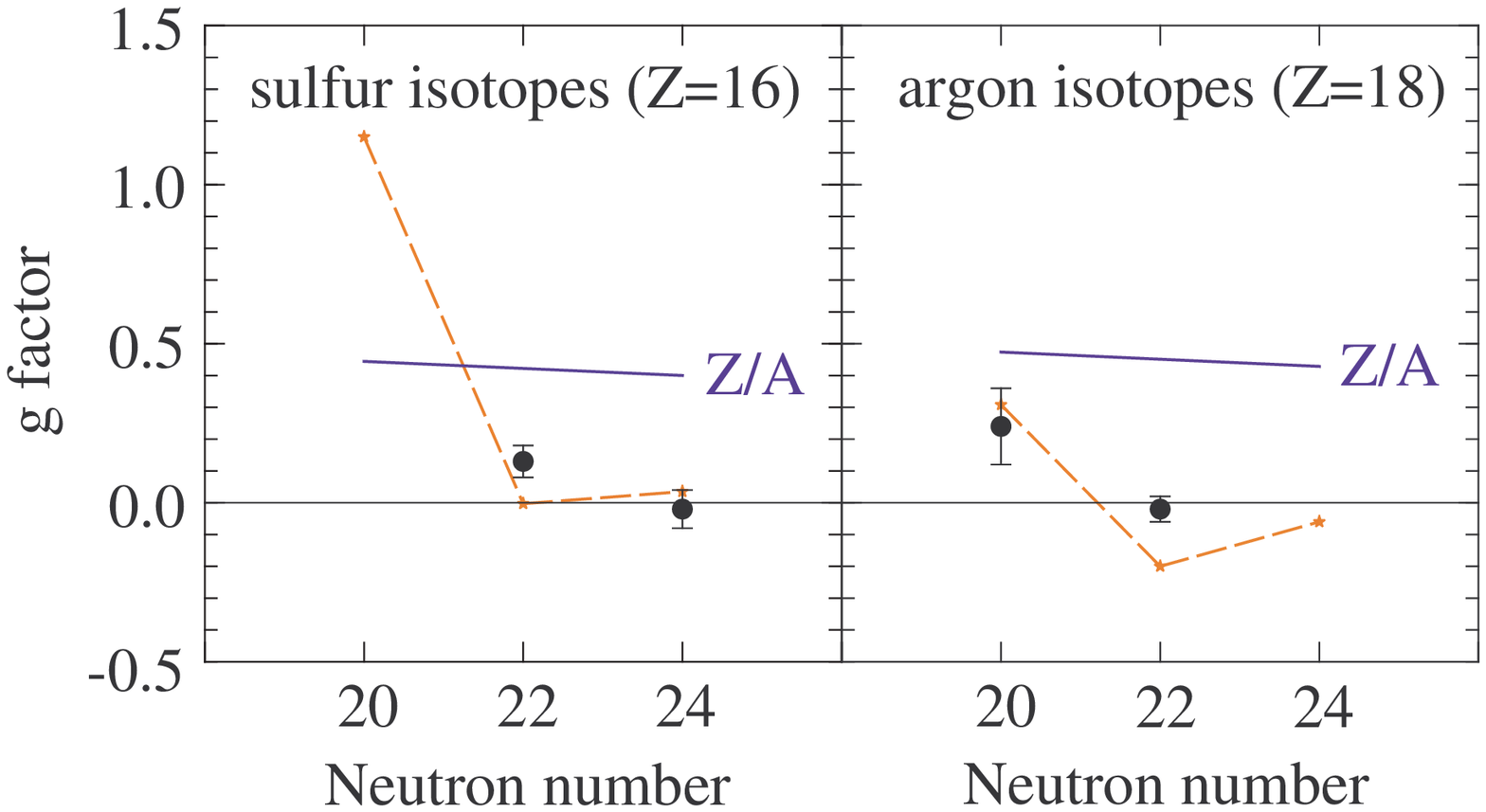}} \caption{(Color
online) Theoretical $g$~factors compared with experiment. The dashed
line shows the present shell model calculations. The previous
results for $^{38}$Ar and $^{40}$Ar are from Refs. \protect
\cite{Spei2005} and \protect \cite{Stef2005}, respectively.}
\label{theoryfig}
\end{center}
\end{figure}

The $g$~factors of the 2$^+_1$ states were evaluated using the bare
nucleon $g$~factors. The calculated $g$~factors are compared with
experimental results in Fig.~\ref{theoryfig} and
Table~\ref{tab:Gtheory}, which also shows the orbital and spin
contributions to the $g$~factor originating from both protons and
neutrons. As will be discussed below, the overall level of agreement
between theory and experiment is satisfactory given the extreme
sensitivity to configuration mixing and the near cancelation of
proton and neutron contributions in the $N=22,24$ isotones.

The main partitions of the 2$^+_1$-state wavefunctions in $^{38}$S
and $^{40}$S are indicated in Table~\ref{tab:wavefunctions}.
Table~\ref{tab:sobd} shows the decomposition of the $g(2^+)$ values
into the contributions from each combination of single-particle
orbitals for $^{38,40}$S and $^{38,40}$Ar. The off-diagonal
contributions from spin-orbit partner orbits can play an important
role. In $^{38}$S and $^{40}$S they tend to quench the (diagonal)
moment of the dominant configuration.

Many authors have argued on experimental and theoretical grounds
that $^{40}$S ($N=24$) is deformed, some linking it to a weakening
of the $N=28$ shell gap (see
Refs.~\cite{Sche1996,Glasm1997,Wing2001,Sohl2002,rod02,cau04,Reta1997}
and references therein). To explore the role of excitations across
the $N=28$ shell gap, a further set of calculations was performed in
which the neutrons were confined to the $f_{7/2}$ orbit. The results
of these calculations, labeled SDF, are given in
Tables~\ref{tab:E2theory} and \ref{tab:Gtheory}.

\begin{table*}
\caption{Theoretical and experimental excitation energies,
quadrupole moments, and $B(E2)$ values. Experimental data are from
\protect \cite{Raman}.}
\begin{ruledtabular}
\begin{tabular}{*{10}c}
Nuclide & Model& $E_{2^+}^{\rm th}$ & $E_{2^+}^{\rm exp}$  &
$Q(2^+_1)$ & $B(E2) \uparrow ^{\rm th}$ & $B(E2) \uparrow ^{\rm
exp}$ &
$\beta^{\rm th}_{Q}$ & $| \beta^{\rm th}_{E2}|$ & $| \beta^{\rm exp}_{E2}|$ \\
& & (MeV) & (MeV) & ($e$ fm$^2$) & ($e^2$ fm$^4$) & ($e^2$ fm$^4$) &\\
\hline
$^{36}$S &SDPF& 3.426 & 3.291 & $-11.6$ & 141 & 104(28) & $+0.21$ &  0.20 & 0.17(2) \\
\\
$^{38}$S &SDPF& 1.531 & 1.292 & $-9.7$ & 268 & 235(30)  & $+0.17$ & 0.26  & 0.25(2) \\
         &SDF & 1.286 &       & $+7.5$ & 158 &          & $-0.13$ & 0.20  &         \\
\\
$^{40}$S &SDPF& 0.980 & 0.904 & $-19.3$ & 473 & 334(36) & $+0.33$ & 0.34 & 0.28(2) \\
         &SDF & 1.052 &  & $-9.8$ & 328 &  & $+0.17$ & 0.28 &  \\
\\
$^{38}$Ar &SDPF& 2.018 & 2.167 & $+4.3$ & 178 & 130(10) & $-0.07$ & 0.19 & 0.16(1) \\
\\
$^{40}$Ar &SDPF& 1.371 & 1.461 & $+8.5$ & 263 & 330(40) & $-0.13$ & 0.22 & 0.25(2) \\
          &SDF & 0.966 &  & $+12.8$ & 228 &  & $-0.20$ & 0.21 & \\
\\
$^{42}$Ar &SDPF& 1.243 & 1.208 & $+5.2$ & 360 & 430(10) & $-0.08$ & 0.25 & 0.28(3) \\
          &SDF & 0.918 &  & $+13.5$ & 317 &  & $-0.20$ & 0.24 &  \\

\end{tabular}
\end{ruledtabular}
\protect \label{tab:E2theory}
\end{table*}

\begin{table*}
\caption{Theoretical $g$~factors compared with experiment. }
\begin{ruledtabular}
\begin{tabular}{*{8}c}
Nuclide & Model& \multicolumn{3}{c}{ $g^{\rm th}_{\rm proton}$ } &
$g^{\rm th}_{\rm neutron}$ & $g^{\rm th}$ & $g^{\rm exp}$
\\ \cline{3-5} & & orbital & spin & total\\
\hline
$^{36}_{16}$S$_{20}$ &SDPF& 0.967 &0.187 & $+1.154$ & 0 & $+1.154$ & \\
\\
$^{38}_{16}$S$_{22}$ &SDPF& 0.225 & 0.073 & $+0.298$ & $-0.301$ &
$-0.003$ & $+0.13(5)$
\footnotemark[1] \\
 &SDF&  0.087 & 0.049 & $+0.136$ & $-0.494$ &
$-0.358$ & \\
\\
$^{40}_{16}$S$_{24}$ &SDPF& 0.225 &0.051 & $+0.276$ & $-0.241$ &
$+0.035$ & $-0.01(6)$
 \footnotemark[1]\\
 &SDF& 0.249 & 0.070 & $+0.318 $ & $-0.404 $
& $-0.085 $ &\\
\\
$^{38}_{18}$Ar$_{20}$ &SDPF& 1.151 &$-0.842$ & $+0.308$ & 0 &
$+0.308$ & $+0.24(12)$
\footnotemark[2]\\
\\
$^{40}_{18}$Ar$_{22}$ &SDPF& 0.311 & $-0.147$ & $+0.164$ & $-0.364$
&
$-0.200$ & $-0.02(4)$ \footnotemark[3]\\
 &SDF& 0.228 & $-0.092 $ & $+0.136 $ & $-0.431 $
& $-0.295$ &\\
\\
$^{42}_{18}$Ar$_{24}$ &SDPF& 0.263 & $-0.043$ & $+0.220$ & $-0.280$ & $-0.060$ &\\
 &SDF& 0.242 & $-0.025$ & $+0.217 $ & $-0.417 $ & $-0.200
$ &\\
\end{tabular}
\footnotetext[1]{Present work.} \footnotetext[2]{Ref. \protect
\cite{Spei2005}} \footnotetext[3]{Ref. \protect \cite{Stef2005}}
\end{ruledtabular}
\protect \label{tab:Gtheory}
\end{table*}

\begin{table*}
\caption{Partitions $ > 5\%$ contributing to the wavefunctions of
the $2_1^+$ states of $^{38}$S and $^{40}$S in the SDPF model.}
\begin{ruledtabular}
\begin{tabular}{*{9}c}
nuclide & \multicolumn{3}{c}{proton orbit} &
\multicolumn{4}{c}{neutron orbit} & occupation (\%) \\ \cline{2-4}
\cline{5-8}
 & $1d_{5/2}$ &$1d_{3/2}$ &$2s_{1/2}$ &$1f_{7/2}$ &$1f_{5/2}$
&$2p_{3/2}$ & $2p_{1/2}$ &\\
\hline
$^{38}$S & 6 & 0 & 2 & 2 & 0 & 0 & 0 & 26.89 \\
& 6 & 1 & 1 & 2 & 0 & 0 & 0 & 18.36 \\
& 6 & 2 & 0 & 2 & 0 & 0 & 0 & 12.10 \\
& 6 & 0 & 2 & 1 & 0 & 1 & 0 & 12.01 \\
& 6 & 1 & 1 & 1 & 0 & 1 & 0 &  6.36 \\
\\
$^{40}$S & 6 & 2 & 0 & 4 & 0 & 0 & 0 & 17.26 \\
& 6 & 1 & 1 & 4 & 0 & 0 & 0 & 16.22 \\
& 6 & 0 & 2 & 4 & 0 & 0 & 0 &  9.30 \\
& 6 & 2 & 0 & 3 & 0 & 1 & 0 &  9.18 \\
& 6 & 1 & 1 & 3 & 0 & 1 & 0 &  8.37 \\
& 6 & 0 & 2 & 3 & 0 & 1 & 0 &  6.05 \\
\end{tabular}
\end{ruledtabular}
\protect \label{tab:wavefunctions}
\end{table*}

\begin{table}
\caption{Contributions to the $g$ factors of the $2_1^+$ states in
$^{38}$S, $^{40}$S, $^{38}$Ar and $^{40}$Ar for the SDPF model.}
\begin{ruledtabular}
\begin{tabular}{crrrr}
\multicolumn{1}{c}{orbits} & \multicolumn{1}{c}{$^{38}$S} &
\multicolumn{1}{c}{$^{40}$S} & \multicolumn{1}{c}{$^{38}$Ar}& \multicolumn{1}{c}{$^{40}$Ar}\\
\hline
$\pi$ $1d_{5/2}$ - $1d_{5/2}$ &   0.046 &   0.067 & 0.044 & 0.033 \\
$\pi$ $1d_{5/2}$ - $1d_{3/2}$ & $-0.015$& $-0.003$& 0.124 & 0.003 \\
$\pi$ $1d_{3/2}$ - $1d_{3/2}$ &   0.014 &   0.014 & 0.080 & 0.021 \\
$\pi$ $2s_{1/2}$ - $2s_{1/2}$ &   0.254 &   0.199 & 0.061 & 0.107 \\
$\nu$ $1f_{7/2}$ - $1f_{7/2}$ & $-0.430$& $-0.447$& & $-0.394$ \\
$\nu$ $1f_{7/2}$ - $1f_{5/2}$ &   0.039 &   0.118 & & $-0.002$ \\
$\nu$ $1f_{5/2}$ - $1f_{5/2}$ &   0.009 &   0.011 & &   0.005  \\
$\nu$ $2p_{3/2}$ - $2p_{3/2}$ &   0.054 &   0.092 & &   0.018  \\
$\nu$ $2p_{3/2}$ - $2p_{1/2}$ &   0.026 & $-0.016$& &   0.008  \\
$\nu$ $2p_{1/2}$ - $2p_{1/2}$ &   0.002 &   0.000 & &   0.001  \\
\end{tabular}
\end{ruledtabular}
\protect \label{tab:sobd}
\end{table}

\subsection{Discussion of results}


In the $N=20$ isotones, $^{36}$S and $^{38}$Ar, the 2$^+_1$ state is
a pure proton excitation for our model space. In $^{36}$S the
first-excited state at 3.29 MeV is dominated (87\%) by the proton
$(s_{1/2} d_{3/2})_{2^+}$ configuration for which $g = +1.459$. The
2$^+_1$ configuration in $^{38}$Ar is predominantly (93\%) $\pi
(d_{3/2})^2_{2^+}$ for which $g = +0.083$. This case demonstrates
the extreme sensitivity of the magnetic moments to configuration
mixing - the remaining 7\% of the wavefunction raises the calculated
moment to $+0.308$, in agreement with the experimental value
\cite{Spei2005}. It also shows the importance of off-diagonal terms
(see Table~\ref{tab:sobd}) as the diagonal contributions associated
with the two main configurations account for only 60\% of the total
theoretical $g$~factor.


Two neutrons have been added to the $pf$ shell in the $N=22$
isotones $^{38}$S and $^{40}$Ar. Since $^{36}$S is almost doubly
magic, the initial expectation might be that the first-excited state
of $^{38}$S would be dominated by the neutron $f_{7/2}$
configuration weakly coupled to the $^{36}$S core, resulting in a
$g$~factor near $-0.3$. An example of this type of behavior is the
$N=52$ isotope $^{92}_{40}$Zr$_{52}$ \cite{Jak1999,Stu2004}; the
effect is also evident, but less pronounced, in
$^{94}_{42}$Mo$_{52}$ \cite{Man2001}. In contrast with this
weak-coupling scenario, the near zero theoretical $g$~factor and the
small but positive experimental $g$~factor of $^{38}$S require
additional proton excitations, which indicate strong coupling
between protons and neutrons. It is relevant to note that strong
proton-neutron coupling is considered one of the prerequisites for
the onset of deformation. The effect on the proton configuration due
to coupling with neutrons in orbits above the $N=28$ shell gap will
be discussed below.


For $N=22,24$ the shell model predicts a cancelation of the proton
and neutron contributions to the moment. Under these conditions, the
description of the $g$~factors is satisfactory. The dependence of
the $g(2^+)$ in $^{40}$Ar on the basis space, the interaction, and
the choice of effective nucleon $g$~factors, has been investigated
in Ref.~\cite{Stef2005}. To estimate the impact of the use of
effective $g$~factors, we recalculated the $g(2^+)$ values in
Table~\ref{tab:Gtheory} adopting the effective $g_l$ and $g_s$
values employed by Stefanova {\em et al}. The effect was to make the
$2^+$-state $g$~factors more positive, by $\sim 0.05$ for the S
isotopes and by $\sim 0.1$ for the Ar isotopes; the discrepancy
between theory and experiment for $^{38}$S and $^{40}$Ar is
therefore reduced, but not eliminated by the use of effective
$g$~factors.

A comparison may also be made between $^{38}$S and $^{40}$S, and
their isotones $^{42}$Ca and $^{44}$Ca, where the experimental
$g$~factors, $g(2^+) = +0.04(6)$ and $g(2^+) = +0.16(3)$, are also
far from that of the $\nu f_{7/2}^2$ configuration
\cite{Sch2003,Tay2003}. In both $^{42}$Ca and $^{44}$Ca there is
evidently a significant collective component in the wavefunction
that cannot readily be included in shell model calculations
\cite{Sch2003,Tay2003}. Some remnant of this collective excitation
may occur in $^{38}$S and $^{40}$S, although it is very much smaller
than in the calcium isotopes.

We conclude that the dominant components of the shell model
wavefunctions are correct and give a satisfactory description of the
measured $g$~factors. As such, the shell model calculations provide
a microscopic explanation of the development of quadrupole
collectivity in these nuclei, which will now be discussed in more
detail.


The existence of deformation in nuclei has long been associated with
strong interactions between a significant number of valence protons
and neutrons, particularly in nuclei near the middle of a major
shell. Without exception the deformed nuclei studied to date have
$g$~factors near the hydrodynamical limit, $Z/A$, reflecting the
strong coupling between protons and neutrons, and a magnetic moment
dominated by the orbital motion of the proton charge with small
contributions from the intrinsic magnetic moments of either the
protons or the neutrons. Examples include $^{24}$Mg in the $sd$
shell \cite{Bro1982}, and $^{50}$Cr in the $pf$ shell
\cite{Ern2000}. Robinson {\em et al.} \cite{Rob2006} have recently
used the shell model to examine the relation between the $B(E2)$,
$Q(2^+_1)$ and the orbital magnetic dipole strength (scissors mode)
in these nuclei.

As noted above, Coulomb-excitation studies and the level scheme of
$^{40}$S suggest that it is deformed. Supporting this
interpretation, the shell model calculations
(Table~\ref{tab:E2theory}) predict consistent intrinsic quadrupole
moments when derived from either the $B(E2)$ or the quadrupole
moment, $Q(2^+_1)$, implying a prolate deformation of $\beta \approx
+0.3$, in agreement with the value deduced from the experimental
$B(E2)$~\cite{Sche1996}. The near zero magnetic moment, however,
does not conform to the usual collective model expectation of $g
\sim Z/A$. Since the shell model calculations reproduce both the
electric and magnetic properties of the 2$^+_1$ state they give
insight into the reasons for this unprecedented magnetic behavior in
an apparently deformed nucleus.

The essential difference between the deformed neutron-rich sulfur
isotopes and the deformed nuclei previously encountered (i.e. either
light nuclei with $N=Z$ or heavier deformed nuclei) is that the spin
contributions to the magnetic moments are relatively more important,
especially for the neutrons. It can be seen from
Table~\ref{tab:Gtheory} that the proton contributions to the
$g$~factors are dominated by the orbital component, as is usually
the case for deformed nuclei, but the substantial neutron
contributions originate entirely with the intrinsic spin. The main
partitions of the shell-model wavefunctions summarized in
Table~\ref{tab:wavefunctions} indicate that both $^{38}$S and
$^{40}$S have a dominant occupation of the neutron $f_{7/2}$ orbit,
for which $g = -0.547$ (unless coupled to spin zero). The net
neutron contribution to the $g$~factor in $^{38}$S ($^{40}$S) is
therefore $\sim 55\%$ ($\sim 44\%$) of that of a pure $f_{7/2}$
neutron configuration. Those configurations with neutron excitations
into the $p_{3/2} f_{5/2} p_{1/2}$ shell mainly have a single
occupation of the $p_{3/2}$ orbit. Although these excitations tend
to reduce the magnetic moment of the neutrons from the value of the
pure $f_{7/2}^n$ configuration, they do not cancel the neutron
magnetic moment altogether because the $g$~factor of the $\nu
(f_{7/2} \otimes p_{3/2})_J$ configuration coupled to spin $J=2$ is
$-0.183$, and the value increases in magnitude for other possible
values of $J$. The off-diagonal terms in the $M1$ operator also
quench the $f_{7/2}$ neutron contribution to $g(2^+)$, as indicated
in Table~\ref{tab:sobd}.

\begin{figure}[btp]
\begin{center}
\resizebox{75mm}{!}{\includegraphics{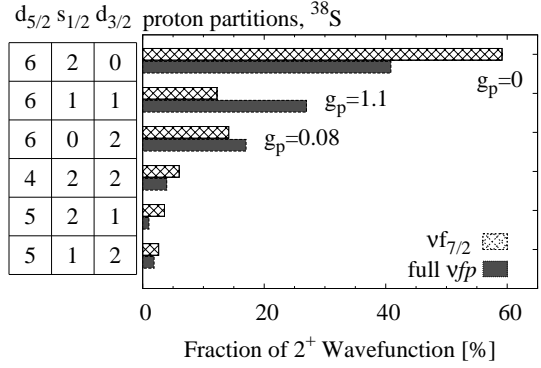}} \caption{Dominant
proton partitions in $^{38}$S comparing calculations in which
neutrons occupy the full $pf$ shell (SDPF) with calculations in
which they are restricted to the $f_{7/2}$ shell (SDF). The
$g$~factors which represent the diagonal contributions of the most
important configurations are indicated.} \label{fig:S38partition}
\end{center}
\end{figure}

\begin{figure}[btp]
\begin{center}
\resizebox{75mm}{!}{\includegraphics{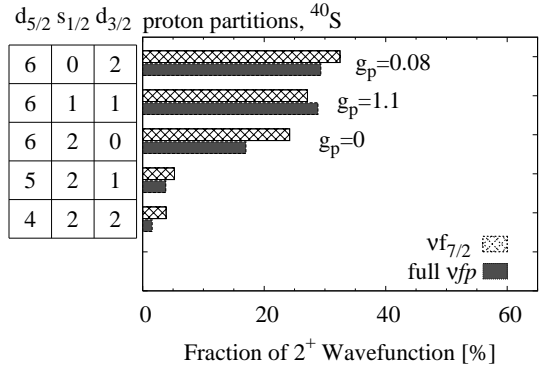}} \caption{As for
Fig.~\protect \ref{fig:S38partition}, but for $^{40}$S.}
\label{fig:S40partition}
\end{center}
\end{figure}

The role of neutron excitations across the $N=28$ shell gap can be
examined further by comparing the calculations using the full SDPF
model space with the truncated SDF calculations in which the
neutrons were restricted to the $f_{7/2}$ shell. Looking first at
the $E2$ properties presented in Table~\ref{tab:E2theory}, it is
apparent from the $Q(2^+)$ values that excitations into the $p_{3/2}
f_{5/2} p_{1/2}$ shell are needed to produce significant prolate
deformations in $^{38}$S and $^{40}$S. Furthermore, in $^{40}$S, the
quadrupole moment and $B(E2)$ are not consistent with the same
intrinsic quadrupole deformation unless the neutrons can occupy the
$p_{3/2} f_{5/2} p_{1/2}$ shell. The wavefunctions in
Table~\ref{tab:wavefunctions} indicate that the development of
quadrupole collectivity depends most sensitively upon the occupation
of the $\nu p_{3/2}$ orbit.

Turning to the $g$~factors, it is apparent that two effects are
important. First, in all cases the neutron contribution to $g(2^+)$
is quenched significantly when the $p_{3/2} f_{5/2} p_{1/2}$ orbits
are occupied. Second, the case of $^{38}$S shows a dramatic change
in the $g$~factor due to a relatively small neutron occupation of
the $p_{3/2}$ orbit, highlighting that there is a strong coupling
between the protons and neutrons. Specifically, a small occupation
of the neutron $p_{3/2}$ orbit, moves the theoretical $g$~factor
from $-0.358$ to $-0.003$, falling short of the experimental value
by a small margin compared with the distance traveled.
Fig.~\ref{fig:S38partition} shows how the proton partition of the
wavefunction for $^{38}$S changes significantly with the occupation
of the $p_{3/2} f_{5/2} p_{1/2}$ shell; there is a significant
increase in the contribution of the proton $s_{1/2} d_{3/2}$
configuration, which has a large $g$~factor, and which has been
proposed to drive deformation. In contrast,
Fig.~\ref{fig:S40partition} shows that for $^{40}$S this proton
configuration contributes strongly to the wavefunction whether
excitations across the $N=28$ gap are allowed or not. Apparently the
effect of a single neutron occupying the $p_{3/2}$ orbit is diluted
when three remain in the $f_{7/2}$ shell.

Since the hydrodynamical collective models fail to explain the
$g$~factor of $^{40}$S, the observed quadrupole collectivity in this
region is better interpreted in terms of the symmetries of the shell
model Hamiltonian. More specifically, the development of quadrupole
collectivity can be linked to the quasi-$SU(3)$ symmetry identified
by Zuker {\em et al.} \cite{Zuk1995} and considered for the sulfur
isotopes by Retamosa {\em et al.} \cite{Reta1997}. In the neutron
space the $\Delta j =2$ orbits $f_{7/2}$-$p_{3/2}$ develop the
quasi-$SU(3)$ symmetry. As shown by the comparisons in
Table~\ref{tab:E2theory}, occupation of the $p_{3/2}$ orbit is
clearly essential if quadrupole collectivity is to emerge. Turning
to the proton space, for the S isotopes of interest the proton
$d_{5/2}$ orbit is essentially closed and only the $d_{3/2}$ and
$s_{1/2}$ orbits are active (see Table~\ref{tab:wavefunctions}).
Although quasi-$SU(3)$ cannot develop for these two orbits, an
approximate pseudo-$SU(3)$ geometry does develop. When neutrons
begin to fill the $pf$ shell, the effective energies of the two
proton orbits become increasingly degenerate, as is manifested by
the narrowing of the gap between the lowest $3/2^+$ and $1/2^+$
states in the odd-$A$ isotopes of K, Cl and P \cite{Cott1998,Fri05}.
In the limit of degenerate $d_{3/2}$ and $s_{1/2}$ orbits the
valence proton space has the geometry of pseudo-$SU(3)$. Within this
framework, quadrupole collectivity develops in the neutron-rich
sulfur isotopes because the proton number is optimal for quadrupole
coherence, despite the fact that the $d_{3/2}$ - $s_{1/2}$
degeneracy is not reached \cite{Reta1997}.

The shift in the effective single particle energies of the proton
$d_{3/2}$ and $s_{1/2}$ orbits is strongly linked to the effect of
the monopole component of the tensor term in the nucleon-nucleon
interaction. Otsuka {\em et al.} \cite{Ots2005} have shown that the
effect of the monopole interaction between the proton $d_{3/2}$ and
neutron $f_{7/2}$ orbits is attractive, which narrows the gap
between the proton $d_{3/2}$ and $s_{1/2}$ states as more and more
neutrons are added to the $f_{7/2}$ shell. This effect of the
monopole interaction is therefore more pronounced in $^{40}$S than
in $^{38}$S, and may explain why the contribution of the proton
partition $d_{5/2}^6 s_{1/2}^1 d_{3/2}^1$ is less sensitive to the
neutron occupation of the $p_{3/2}$ orbit in $^{40}$S than in
$^{38}$S.

\section{Summary and conclusion}

In summary, we have developed a high-velocity transient-field
technique to measure the $g$~factors of excited states of
neutron-rich nuclei produced as fast radioactive beams. The
$g$~factors of the first-excited states in the neutron-rich isotopes
$^{38}$S and $^{40}$S have been measured with sufficient precision
to test and challenge shell model calculations.

Keys to the success of these measurements on low intensity
radioactive beams (10$^4$ to 10$^5$ pps) include (i) the use of
intermediate energy Coulomb-excitation to align the nuclear spin,
and (ii) exploitation of the high velocities of the fragment beams
to maximize the transient-field precession angle. In these
measurements the precession angle per unit $g$~factor is $\sim 10$
times that in conventional measurements on neighboring nuclei
\cite{Spei2005,Stef2005,Sch2003,Tay2003}.

The nature and origin of the quadrupole collectivity that develops
between $N=20$ and $N=28$, including the role of excitations across
the $N=28$ shell gap, has been explored by comparing the measured
$g$~factors with shell model calculations.

The case of $^{38}$S highlights the role of strong proton-neutron
interactions, showing that neutron excitations across the $N=28$
shell gap cause a large change in the proton configuration. Taken
together with the results for $^{40}$S, it seems that excitations
across the $N=28$ shell gap {\em and} the reduction in the proton
$s_{1/2}$-$d_{3/2}$ gap due to increasing neutron occupation of the
$f_{7/2}$ orbit {\em both} contribute to the development of
collectivity.

Since the hydrodynamical collective models fail to explain the
$g$~factor of $^{40}$S, the observed quadrupole collectivity in this
region should be interpreted in terms of the symmetries of the shell
model Hamiltonian. The unusual combination of magnetic and electric
properties in $^{40}$S apparently comes about because relatively few
particles are involved in the collective motion. These features
reemphasize the unique mesoscopic nature of the nucleus.

Looking to future applications of the high-velocity transient-field
technique, it may be noted that the 2$^+_1$ state in the $N=20$
nucleus $^{32}_{12}$Mg$_{20}$ has a similar excitation energy,
lifetime and $B(E2)$  to $^{40}$S. However the $g$~factor in
$^{32}$Mg might be closer to that of a conventional collective
nucleus since the $N=20$ shell closure is known to vanish far from
stability. Indeed a shell model calculation using the interactions
of Warburton, Becker and Brown \cite{War1990} gives $g = g_p + g_n =
0.52 - 0.11 = +0.41$, close to the collective value of $Z/A=0.375$.

\section*{Acknowledgments}

We thank the NSCL operations staff for providing the primary and
secondary beams for the experiment. We are grateful to Professor N.
Benczer-Koller (Rutgers University) for performing the magnetometer
measurement. This work was supported by NSF grants PHY-01-10253,
PHY-99-83810, PHY-02-44453, and PHY-05-55366. AES, ANW, and PMD
acknowledge travel support from the ANSTO AMRF scheme (Australia).

\bibliography{AESsulfurHVTF}

\end{document}